\documentclass[12pt]{article}
\usepackage{epsfig}
\font\myit=cmti10

\newcommand{\be}{\begin{equation}}
\newcommand{\ee}{\end{equation}}
\newcommand{\beq}{\begin{eqnarray}}
\newcommand{\eeq}{\end{eqnarray}}

\begin{document}

\title{Signals of non-extensive statistical mechanics in high-energy nuclear 
collisions}

\author{W.M. Alberico$^{1,2}$, P. Czerski$^{3}$, A. Lavagno$^{4,2}$, 
M. Nardi$^{5,2}$, V. Som\'a$^1$
\vspace*{0.3cm}\\
 {\myit        $^1$Dipartimento di Fisica Teorica, Universit\`a di Torino,}\\
   {\myit         Via P. Giuria 1, I-10126 Torino, Italy} \\
   {\myit     $^2$Istituto Nazionale di Fisica Nucleare, Sezione di Torino}\\
   {\myit     $^3$The Henryk Niewodniczanski Institute of Nuclear Physics,}\\
   {\myit        Polish Academy of Sciences, Krakow, Poland}\\
   {\myit     $^4$Dipartimento di Fisica, Politecnico di Torino,}\\
   {\myit       C.so Duca degli Abruzzi 24, I-10129 Torino, Italy} \\
  {\myit  $^5$Centro Studi e Ricerche E.Fermi, Compendio Viminale 00184 Roma}\\
          }
\maketitle

\begin{abstract}

Starting from the presence of non-ideal plasma effects due to strongly coupled 
plasma in the early stage of relativistic heavy-ion collisions, we 
investigate, from a phenomenological point of view, the relevance of 
non-conventional 
statistical mechanics effects on the rapidity spectra of net 
proton yield at AGS, SPS and RHIC. We show that the broad rapidity shape 
measured at RHIC can be very well reproduced in the framework of a non-linear 
relativistic Fokker-Planck equation which incorporates non-extensive 
statistics and anomalous diffusion.

\end{abstract}

\section{Introduction}

It is expected that hadrons dissociate into a plasma of
their elementary constituents, quarks and gluons (QGP), at a density
several times the nuclear matter density and/or  at temperatures of the
order of $T_c=170$ MeV, which is the critical temperature of the 
transition from the hadronic gas phase to the QGP phase, as expected 
from lattice QCD calculations.
In central Au+Au collisions at RHIC energy densities are reached
that are far above the critical energy density (of the order of 1
GeV/fm$^3$) for a transition to the QGP phase.

Since interactions among quarks and gluons become weak at small
distance or high energy, one usually expects that QGP is a weakly
interacting, ideal plasma, which can be described by perturbative QCD. 
However, this is rigorously true only at very high temperature, while 
non-perturbative phenomena prevail up to temperatures of several times  
$T_c$~\cite{karsch}.  For example, as it is known from several approaches,
partons in a hot plasma acquire a temperature dependent Debye mass, which
deeply modifies the QCD dynamics, yet inducing collective effects. 
Hence the final states in an ultrarelativistic nucleus-nucleus collision 
can be very sensibly affected by non-ideal plasma effects, including 
long-range interactions and memory effects. 
Many properties concerning the formation of the expected new phase 
and the consequent hadronization process are still under 
debate~\cite{albe,peshier,thoma1,schmidt}.
Pre-thermalization or metastability conditions can be realized and
the standard equilibrium statistical mechanics assumptions can not
be taken for granted in the description of the system toward 
equilibrium~\cite{berges}.

In this paper, we want to investigate the relevance of such non-ideal and
non-equilibrium plasma effects within a phenomenological study of
the net-baryon rapidity distribution  in various relativistic heavy ion 
collisions, up to the central Au+Au collisions at 
the highest RHIC energy of $\sqrt{s_{_{NN}}}$=200 GeV. The
interest of this observable lies in the fact that in these processes 
 nuclear matter reaches high energy densities and nuclei undergoing 
central collisions strongly reduce their original longitudinal momentum. 
This loss of rapidity, usually referred to as baryon stopping, is an important
characteristic to understand the reaction mechanism at high energy
density and the net rapidity distribution is very sensitive to the
dynamical and statistical properties of nucleus-nucleus at high energy.

We study the evolution of the rapidity distribution from a
macroscopic point of view by using a non-linear relativistic
Fokker-Planck equation and we will show that the observed broad
rapidity shape could be a signal of non-equilibrium properties of
the system in which non-extensive Tsallis statistical mechanics
\cite{tsallis,gellmann} emerges in a natural way.
 Similar approaches have been considered in the past to analyze 
transverse momentum distributions~\cite{albe,biro04,biro2,biro5,wilk} 
and power law spectra at large $p_\perp$ in terms of various non conventional 
extensions of the Boltzmann-Gibbs thermostatistics, including the Tsallis one.

The net-proton rapidity distribution appears to be a more elusive 
observable with respect to equilibrium considerations, since the main 
component is expected to be related to the initial beam protons; however
the results obtained at RHIC do not appear easily reproduced within 
standard approaches. Encouraging results have been obtained by 
Wolschin~\cite{wol1,wol2,wol3} within a three-components Relativistic 
Diffusion Model (RDM). Here again a possible link with Tsallis 
statistics is suggested~\cite{wol2}.

The paper is structured as follows. In Sec.~2, we qualitatively
review some basic features of a non-ideal QGP like the one hypothetically
generated in the relativistic heavy-ion collisions at AGS, SPS and
RHIC energies. We also show that the main assumptions contained in
the derivation of the standard dynamical kinetic equations,
describing the evolution of the system toward the equilibrium, are
no longer valid. In Sec.~3, we introduce a non-linear relativistic
kinetic equation containing anomalous diffusion effects 
in the framework of the non-extensive Tsallis thermostatistics. In Sec.~4, we
study the net proton rapidity spectra comparing the AGS, SPS and
RHIC data. Conclusions are reported in Sec.~5.

\section{Non-ideal QGP plasmas and kinetic assumptions towards the equilibrium}

In the literature, an ordinary plasma is usually characterized by the 
value of the plasma parameter $\Gamma$ \cite{ichi} 
\be 
\Gamma=\frac{\langle
U\rangle}{\langle T \rangle} \; , 
\ee 
defined as the ratio between  potential
energy $\langle U\rangle$ versus kinetic energy $\langle T \rangle$. 
When $\Gamma\ll 1$, one has a dilute weakly
interacting gas; the Debye screening length $\lambda_D$ is much
greater than the average interparticle distance $r_0$ 
and a large number of particles is contained in the
Debye sphere. Binary collisions induced by screened forces produce,
in the classical case, the standard Maxwell-Boltzmann velocity
distribution. If $\Gamma\approx 0.1\div 1$, then
$\lambda_D\approx r_0$, and it is not possible to clearly separate
individual and collective degrees of freedom: this situation refers to 
a weakly interacting, non-ideal plasma. 
Finally, if $\Gamma\geq 1$, the plasma is
strongly interacting, Coulomb interaction and quantum effects
dominate and determine the structure of the system.

The quark-gluon plasma close to the critical temperature is a
strongly interacting system. In fact, following 
Ref.\cite{albe,peshier,thoma1}, the
color-Coulomb coupling parameter of the QGP is defined, in analogy
with the one of the classical plasma, as 
\beq 
\Gamma \approx  C \frac{g^2}{4\pi r_0\, T} \; ,
\eeq 
where $C=4/3$ or 3 is the Casimir invariant for the quarks or gluons, 
respectively;  for typical temperatures attained in relativistic heavy
ion collisions, $T\simeq 200$~MeV, $\alpha_s=g^2/(4\pi)=0.2 \div 0.5$, and 
$r_0\simeq n^{-1/3}\simeq 0.5$~fm ($n$ being the particle density for an
ideal gas of 2 quark flavors in QGP). Consequently, one obtains 
$\Gamma \simeq 1.5-5$ and the plasma can be considered to be in a non-ideal 
 liquid phase~\cite{peshier,thoma1}.

In these conditions, the generated QGP does not satisfy anymore
the basic assumptions (BBGKY hierarchy) of a kinetic equation
(Boltzmann or Fokker-Planck equation) which describes a system
toward the equilibrium. In fact, near the phase transition the
interaction range is much larger than the Debye screening length and
a small number of partons is contained in the Debye sphere~\cite{albe,thoma1}.
Therefore, the collision time is not much smaller than the mean
time between collisions and the interaction is not local. The binary
collisions approximation is not satisfied, memory effects and
long--range color interactions give rise to the presence of
non--Markovian processes in the kinetic equation, thus  affecting the
thermalization process toward equilibrium as well as the standard
equilibrium distribution. In the next section we will see how such effects can
be taken into account, from a macroscopic point of view,
by means of an appropriate generalization of the standard
statistical mechanics.

\section{Relativistic non-extensive kinetic equations}

In many-body long-range-interacting systems, it has been recently
observed the emergence of long-living quasi stationary
(metastable) states characterized by non-Gaussian power law
velocity distributions, before the Boltzmann-Gibbs equilibrium is
attained.  In this case, the standard statistical mechanics is no longer 
appropriate to describe such a behavior~\cite{latora,borges,ananos}. 
Indeed, the basic assumption of the Boltzmann-Gibbs 
statistical mechanics is that the system can be subdivided into a set 
of non-overlapping subsystems,  the total entropy of which is the sum
of the entropies of the independent subsystems.
In the presence of memory effects and long range 
forces the entropy, which is a measure of the information about the 
particle distribution in the states available to the system, does not 
satisfy the extensivity property. 

Recently, there is an increasing evidence that the generalized
non-extensive statistical mechanics, proposed by Tsallis
\cite{tsallis,gellmann}, can be considered as an appropriate
basis for a theoretical framework to describe physical phenomena
where long-range interactions, long-range microscopic memories
and/or fractal space-time constraints are present. A considerable
variety of physical issues show a quantitative agreement
between experimental data and theoretical models based on Tsallis'
thermostatistics. In particular, there is a growing interest to
high energy physics applications of non-extensive statistics.
Several authors outline the possibility that experimental
observations in relativistic heavy-ion collisions can reflect
non-extensive statistical mechanics effects during the early stage of 
the collisions and the thermalization evolution of the system
\cite{albe,biro2,wilk,lava2,rafe,biro1,biro3}.

In order to study from a phenomenological point of view 
experimental observables in relativistic heavy-ion collisions, we
can introduce the basic macroscopic variables in the language of
relativistic kinetic theory following the Tsallis' prescriptions
for the non-extensive statistical mechanics. In this framework the
particle (four-vector) flow can be generalized as~\cite{lava}
\begin{equation}
N^\mu(x)=\frac{1}{Z_q}\int \frac{d^3p}{p^0} \, p^\mu \,f(x,p) \; ,
\label{nmu}
\end{equation}
and the energy-momentum flow as
\begin{equation}
T^{\mu\nu}(x)=\frac{1}{Z_q}\int \frac{d^3p}{p^0} \, p^\mu p^\nu \,
\left[f(x,p)\right]^q \; , \label{tmunu}
\end{equation}
where we have set $\hbar=c=1$, $x\equiv x^\mu=(t,{\bf x})$,
$p\equiv p^\mu=(p^0,{\bf p})$ and $p^0=\sqrt{{\bf p}^2+m^2}$ is
the relativistic energy. In the above 
${Z_q}=\int d\Omega \, [f(x,p)]^q $ is the non-extensive partition 
function, $d\Omega$ stands for the corresponding phase space volume element 
and $q$ is the deformation parameter. 
The limit  $q\rightarrow 1$ corresponds to 
ordinary statistics. The four-vector $N^\mu=(n,{\bf j})$
contains the probability density $n=n(x)$ (which is normalized to
unity) and the probability flow ${\bf j}={\bf j}(x)$. The
energy-momentum tensor contains the normalized $q$-mean
expectation value\footnote{The $q$-mean expectation value is defined 
as~\cite{gellmann,lava}:
\begin{equation}
\langle O(x)\rangle_q =\int d^3p \, O(x,p) [f(x,p)]^q\Big{/}
\int d\Omega [f(x,p)]^q \; .
\end{equation}} 
of the energy density, as well as the energy
flow, the momentum and the momentum flow per particle.

On the basis of the above definitions, one can show that
it is possible to obtain a generalized non-linear relativistic
Boltzmann equation \cite{lava}
\begin{equation}
p^\mu \partial_{\mu}\left[f(x,p)\right]^q=C_q(x,p)  \; , \label{boltz}
\end{equation}
where the function $C_q(x,p)$  implicitly defines a generalized
non-extensive collision term
\begin{eqnarray}
C_q(x,p)=&&\frac{1}{2}\! \int\!\!\frac{d^3p_1}{p^0_1}
\frac{d^3p{'}}{p{'}^0} \frac{d^3p{'}_1}{p{'}^0_1} \Big
\{h_q[f{'},f{'}_1]  W(p{'},p{'}_1\vert p,p_1) \nonumber \\
&&-h_q[f,f_1]  W(p,p_1\vert p{'},p{'}_1) \Big\}\; .
\end{eqnarray}
Here $W(p,p_1\vert p{'},p{'}_1)$ is the transition rate between a
two-particle state with initial four-momenta $p$ and $p_1$ and a
final state with four-momenta $p{'}$ and $p{'}_1$; $h_q[f,f_1]$
is the $q$-correlation function relative to two particles in the same
space-time position but with different four-momenta $p$ and $p_1$,
respectively. Such a transport equation conserves the
probability normalization (number of particles) and is
consistent with the energy-momentum conservation laws. The
collision term contains a generalized expression of the molecular
chaos and for $q>0$ implies the validity of a generalized
$H$-theorem, if the following, non-extensive, local four-density
entropy is assumed 
%
\begin{equation}
S_q^\mu(x)=-k_{_B} \,\int \frac{d^3p}{p^0} \,p^\mu f[(x,p)]^q 
[\ln_q f(x,p)-1] \; , \label{entro4}
\end{equation}
where we have used the definition $\ln_q x=(x^{1-q}-1)/(1-q)$.
At equilibrium, the solution of the above
Boltzmann equation is a relativistic Tsallis-like (power law)
distribution and can be written as
\begin{equation}
f^{eq}(p)= \frac{1}{Z_q}\left [1-(1-q) \frac{p^\mu U_\mu}{k_{_B}T}
\right]^{1/(1-q)} \; .
\label{tsarel}
\end{equation}
In the limit $q\rightarrow 1$, Eq.(\ref{tsarel}) reduces to the standard 
relativistic equilibrium J\"uttner distribution.

The above relations represent the basic framework in which, in the
next section, will be studied the net-baryon rapidity distribution
near  equilibrium.

\section{Net-baryon rapidity distribution}

The energy loss of colliding nuclei is a fundamental quantity in 
order to determine the energy available for particle production in 
heavy-ion collisions. Since the baryon number is conserved and
rapidity distributions are only affected by rescattering in the late
stages of the collision, the measured net-baryon ($B-\overline{B}$)
distribution retains information about the energy loss and allows one
to obtain the degree of nuclear stopping.

Recent results for net-proton rapidity spectra in central Au+Au
collisions at RHIC \cite{brahms} show an unexpectedly large
rapidity density at midrapidity in comparison with analogous
spectra at lower energy at SPS \cite{na49} and AGS \cite{ags}. As
outlined from different authors, such spectra can reflect
non-equilibrium effects even if the energy dependence of the
rapidity spectra is not very well understood \cite{brahms,giapu}.

We are going to study the evolution of the rapidity distribution
from a macroscopic point of view by means of a non-linear
relativistic Fokker-Planck equation which can be viewed as the
near-equilibrium approximation of the generalized non-linear
Boltzmann equation (\ref{boltz}). 
 Therefore, the use of the Fokker-Planck equation appears 
to be correct only not too far from equilibrium
\cite{wol1,wol2,wol3,biro1,wilk03,mustafa,rapp,reigert,brantov}.

In order to study the rapidity spectra, it is convenient to separate
the kinetic variables into their transverse and longitudinal components,
the latter being related to the rapidity $y$. If we assume that the 
particle distribution function $f(y,m_\perp,t)$, at fixed transverse mass 
$m_\perp=\sqrt{m^2+p_\perp^2}$, 
is not appreciably influenced by the transverse dynamics (which is
considered in thermal equilibrium), the non-linear Fokker-Planck
equation in the rapidity space $y$ can be written as
\begin{equation}
\frac{\partial}{\partial
t}[f(y,m_\perp,t)]=\frac{\partial}{\partial y} \left [
J(y,m_\perp) [f(y,m_\perp,t)]+D \frac{\partial}{\partial
y}[f(y,m_\perp,t)]^\mu \right ] \label{nlfpe} \; ,
\end{equation}
where $D$ and $J$ are the diffusion and drift coefficients,
respectively. 

Tsallis and Bukman \cite{tsabu} have shown that, for
linear drift, the time dependent solution of the above equation
is a Tsallis (non-relativistic) distribution with $\mu=2-q$ and that 
a value of $q\ne 1$ implies anomalous diffusion, 
i.e., $[y(t)-y_M(t)]^2$ scales like $t^\alpha$, with $\alpha=2/(3-q)$. 
For $q<1$, the above equation implies anomalous subdiffusion, while for $q>1$, 
we have a superdiffusion process in the rapidity space.

Let us note that a similar approach, within a linear and non-linear 
Fokker-Planck equation, has been previously studied in 
Ref.\cite{wol1,wol2,wol3} involving  directly the time evolution of the 
rapidity distribution instead of the particle 
distribution function $f(y,m_\perp,t)$, as in the Eq.(\ref{nlfpe}). 
The two approaches are equivalent only if the particle distribution is 
completely 
decoupled in the transverse and in the longitudinal coordinates and the 
drift coefficient $J$ does not depend on transverse momentum. 
This is the case, for example, when the drift coefficient is assumed to be 
linear in the rapidity space. 
In Ref.\cite{wol1,wol2,wol3}, the author first study a linear Fokker-Planck 
equation with a linear drift coefficient and a free
parameter diffusion coefficient (connected to a collective longitudinal 
expansion)  which implies a strong violation of the fluctuation-dissipation 
theorem. This approach fails to reproduce the rapidity spectra at RHIC energy 
unless it is assumed that a fraction of net protons, near midrapidity, 
undergoes a fast transition to local thermal equilibrium. The non-linear 
Fokker-Planck equation for the rapidity distribution function 
is also considered in Ref.\cite{wol1,wol2,wol3}, even if only approximate 
solutions are considered by means of a linear superposition of power-law 
distribution functions.

We claim that the choice of the diffusion and the drift coefficients plays a 
crucial r\^ole in the solution of the above non-linear Fokker-Planck 
equation (\ref{nlfpe}). 
Such a choice influences the time evolution of the system and 
its equilibrium distribution. 
 This important point has been extensively studied in Ref.\cite{biro04} 
where the authors obtained a generalized fluctuation-dissipation theorem, 
which implies an implicit relation between the drift and diffusion 
coefficient in the framework 
of a linear non-relativistic Fokker-Planck equation.

We are going to show that by generalizing the Brownian motion in a
relativistic framework, the standard Einstein relation is
satisfied and Tsallis non-extensive statistics emerges in a
natural way from the non-linearity of the Fokker-Planck equation.
In fact, by imposing the validity of the Einstein relation for
Brownian particles and setting henceforward the Boltzmann constant $k_{_B}$ 
to unity, we can generalize to the relativistic case the
standard expressions of diffusion and drift coefficients as
follows
\begin{equation}
D=\gamma \, T\; , \ \ \ \  \ J(y,m_\perp)=\gamma \, m_\perp \sinh(y)
\equiv  \gamma \, p_\parallel \; , \label{coeffi}
\end{equation}
where $p_\parallel$ is the longitudinal momentum, $T$ is the
temperature and $\gamma$ is a common constant. 
Let us remark that the above definition of the diffusion and drift
 coefficients,  previously introduced by us in Ref.\cite{lava2}, appears 
as the natural generalization to the relativistic Brownian case in the 
rapidity space. The drift coefficient which 
is, as usual, {\it linear} in the longitudinal momentum $p_\parallel$ becomes 
{\it non-linear} in the rapidity coordinate.

It is easy to see
that the above coefficients give us the Boltzmann stationary
distribution in the linear case ($q=\mu=1$), while the equilibrium 
solution $f^{eq}(y,m_\perp)$ of Eq.(\ref{nlfpe}), with $\mu=2-q$, is a
Tsallis-like distribution (introduced in Eq.(\ref{tsarel})) 
with the relativistic energy $E=m_\perp \cosh(y)$
\begin{equation}
f^{eq}(y,m_\perp)= \Big [1-(1-q)\, m_\perp \cosh(y)/T
\Big ]^{1/(1-q)} \; .
\end{equation}
Out of equilibrium the rapidity distribution at fixed time can be
obtained by means of numerical integration of Eq.(\ref{nlfpe}) with
 delta function initial conditions depending upon the value of the
experimental projectile rapidities. The rapidity distribution at
fixed time is then obtained by numerical integration over the
transverse mass $m_\perp$ as follows
\begin{equation}
\frac{dN}{dy}(y,t)=c\, \int_m^\infty \!\! m^2_\perp \, \cosh(y) \,
f(y,m_\perp,t) \, dm_\perp \; , \label{raint}
\end{equation}
where $m$ is the mass of the considered particles and 
$c$ is the normalization constant, fixed by the experimental data.
The rapidity spectra calculated from Eq.(\ref{raint})
will ultimately depend on two parameters: the ``interaction'' time
$\tau_{int}=\gamma t$ and the non-extensive parameter $q$.

Let us observe that in the numerical solution of Eq.(13) we have not 
explicitly assumed the presence of longitudinal flow. This fact does not 
exclude the dynamical effects of a collective (mainly due to rescattering) 
flow but rather incorporates a description of it in the adopted non-extensive 
statistical mechanics. In fact, as explicitly shown in Ref.~\cite{albe} by 
studying the transverse mass spectrum, dynamical collective interactions are 
intrinsically involved in the generalized statistical mechanics and,  
in a purely thermal source, a generalized $q$-blue shift factor (depending 
on $q$ and on $m_\perp$) appears. We expect a similar behavior also in the 
longitudinal degrees of freedom of the system. 
Let us also notice that collective transverse flow effects 
in the framework of a non-extensive statistical mechanics have been 
investigated in Ref.~\cite{biro04} as well.

In Figs.~1 and 2 we plot the numerical solution of Eq.s (\ref{nlfpe}) and 
(\ref{raint}) at different interaction times $\tau_{int}$ in the linear case 
($q=1$) and in the non-linear case, with $q=1.25$, keeping the same 
initial conditions. 
In comparing Fig.~1 with Fig.~2, we can observe that the rapidity spectra 
appear to be broader in the non-linear case.
 This is a consequence of the anomalous (super)diffusion process 
 implied in the time evolution of Eq.(\ref{nlfpe}) 
and in the nature of the power law distribution 
function which, for $q>1$, enhances the probability to find particles at 
high rapidity values.

\begin{figure}[htb]
\mbox{\epsfig{file=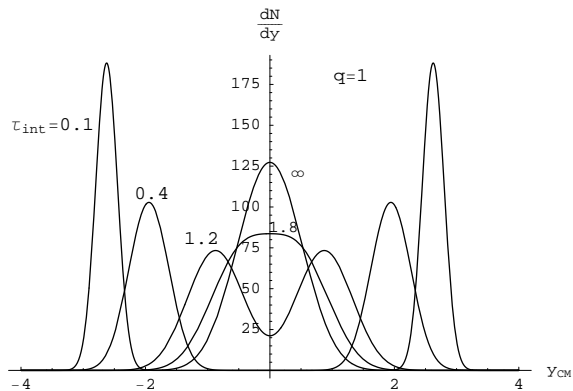,width=0.8\textwidth}}
\vspace{-8.5cm} \caption[]{Numerical solution of Eq.s(\ref{nlfpe}) and 
(\ref{raint}) at different interaction times $\tau_{int}$ in the linear case 
($q=1$).}
\end{figure}

\begin{figure}[htb]
\mbox{\epsfig{file=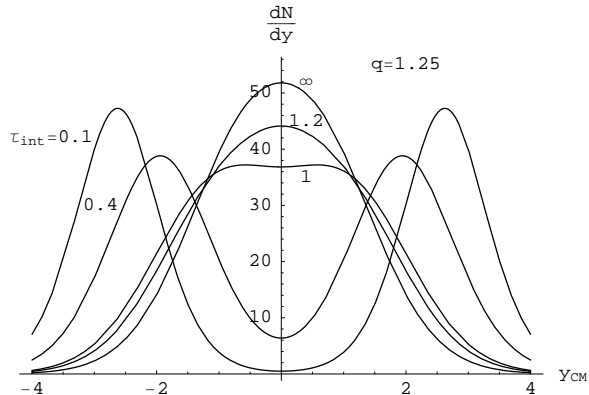,width=0.8\textwidth}}
\vspace{-8.5cm} \caption[]{ The same of Fig.1 but in the
non-linear case with $q=1.25$.}
\end{figure}

In Fig.~3, we report the obtained rapidity distribution (full
line) for the net proton production ($p-\overline{p}$) compared 
with the experimental data of RHIC (Au+Au at $\sqrt{s_{NN}}=200$ GeV,
\cite{brahms}), SPS (Pb+Pb at $\sqrt{s_{NN}}=17.3$ GeV,
\cite{na49}) and AGS (Au+Au at $\sqrt{s_{NN}}=5$ GeV, \cite{ags}). 
 The parameters employed for the three curves are: $q=1.485$ with 
$\tau_{int}=0.47$ for RHIC, $q=1.235$ with $\tau_{int}=0.84$ for SPS and 
$q=1.09$ with $\tau_{int}=0.95$ for AGS, respectively.
We notice that, although $q$ and $\tau_{int}$ appear, in principle, as 
independent parameters, in fitting the data they are not.
Indeed, we can see that only in the non-linear case ($q\ne 1$) there 
exists one and only one (finite) time  $\tau_{int}$ for which the obtained 
rapidity spectrum well reproduces the broad experimental shape. On the 
contrary, for $q=1$, no value of  $\tau_{int}$ can be found, which allows 
to reproduce the data.

We obtain a remarkable agreement with the experimental data by increasing 
the value of the non-linear deformation parameter $q$ as the beam energy 
increases. At AGS energy, the non-extensive statistical effects are 
negligible and the spectrum is well reproduced within 
the standard quasi-equilibrium linear approach. 
At SPS energy, non-equilibrium effects and non-linear evolution become 
remarkable ($q=1.235$) and such effects are even more evident 
for the very broad RHIC spectra ($q=1.485$). 
Let us observe that such an excellent agreement with the RHIC experimental 
data  has not been reached in the similar (but different, as pointed out 
in the introduction) 
approach of Ref.\cite{wol1,wol2,wol3}, especially in reproducing the
experimental points far from the midrapidity region.

From a phenomenological point of view, we can read the larger value of the 
parameter $q$ for the RHIC data, corresponding to non-linear anomalous 
(super)diffusion, as a signal of the non-ideal nature 
of the plasma formed at a temperature larger than the critical one. 
As confirmed by recent microscopic calculations~\cite{peshier,thoma1}, 
strongly coupled non-ideal plasma is generated at energy densities 
corresponding to the order of the critical phase transition temperature 
and in such a regime we find, in our macroscopic approach, 
strong deviations from the standard thermostatistics. 
At much higher energy, such as LHC, we can expect a minor relevance of such 
non-ideal effects since the considerable energy density reached is 
far above the critical one. However, an anomalous diffusion behavior should 
still affect the time evolution  process toward the equilibrium due to 
long range color magnetic forces which remain unscreened (in leading order) 
at all temperatures.

\begin{figure}[htb]
\mbox{\epsfig{file=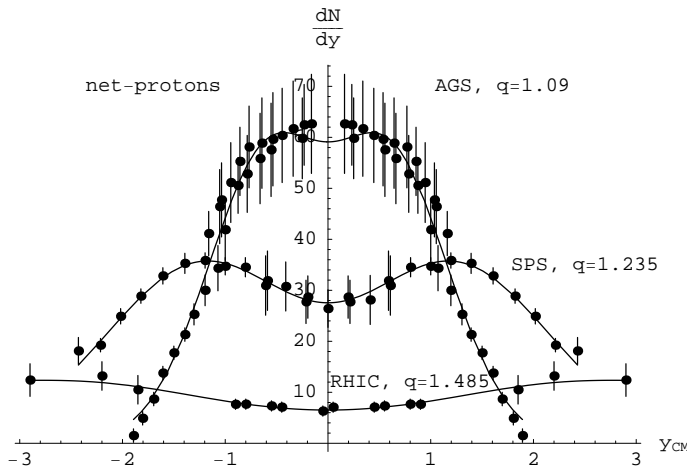,width=0.85\textwidth}}
\vspace{-8.5cm} \caption[]{Rapidity spectra for net proton
production ($p-\overline{p}$) at RHIC (Au+Au at $\sqrt{s_{NN}}=200$
GeV, BRAHMS data), SPS (Pb+Pb at $\sqrt{s_{NN}}=17.3$ GeV, NA49
data) and AGS (Au+Au at $\sqrt{s_{NN}}=5$ GeV, E802, E877, E917). }
\end{figure}

\section{Conclusions}

We have studied the rapidity spectra from a macroscopic point of
view by means of a non-linear kinetic equation which preserves 
the fluctuation-dissipation theorem by introducing an appropriate relativistic 
generalization of the drift and diffusion coefficients. Such a generalized  
evolution equation lies inside the framework of Tsallis' non-extensive 
thermostatistics and contains anomalous diffusion effects which are strongly 
related to the presence of 
non-Markovian memory interactions and long-range color forces.
Multiparticle rescattering, very large at SPS and at RHIC, could be a signal
 of this effects. We recall that 
non-extensive features in relativistic collisions are suggested in
several works, even if a microscopic justification of these
effects is still lacking and lies out of the scope of this paper. 
 The relevant point which was found in the present work is that the larger 
the beam energy of the heavy ions, 
the greater are the values of the deformation parameter $q$, which are 
required to reproduce the experimental rapidity spectra. 
This behavior of the $q$-parameter can be viewed 
as a phenomenological indication of the onset of a strongly coupled 
QGP plasma phase in the early stages of the collision.

\end{document}